# Mechanism of enhancement of electromagnetic properties of MgB$_2$ by nano-SiC doping

S.X. Dou[1*], O. Shcherbakova[1], W.K. Yeoh[1], J.H. Kim[1], S. Soltanian[1], X.L. Wang[1], C. Senatore[2], R. Flukiger[2], M. Dhalle[3], O. Husnjak[4], and E. Babic[4]

[1]Institute for Superconducting and Electronic Materials, University of Wollongong, Wollongong, NSW 2522 Australia
[2]Department of Physics, University of Geneva, Geneva, Switzerland
[3]Faculty of Science and Technology, University of Twente, 7500 AE Enschede, The Netherlands
[4]Department of Physics, Faculty of Science, University of Zagreb, Zagreb, Croatia

A comparative study of pure, SiC, and C doped MgB$_2$ wires has revealed that the SiC doping allowed C substitution and MgB$_2$ formation to take place simultaneously at low temperatures. C substitution enhances $H_{c2}$, while the defects, small grain size and nanoinclusions induced by C incorporation and low temperature processing are responsible for the improvement in $J_c$. The irreversibility field ($H_{irr}$) for the SiC doped sample reached the benchmarking value of 10 T at 20 K, exceeding that of NbTi at 4.2 K. This dual reaction model also enables us to predict desirable dopants for enhancing the performance properties of MgB$_2$.

The significant enhancement of critical current density ($J_c$), irreversibility field ($H_{irr}$), and upper critical field ($H_{c2}$) in MgB$_2$ by nano-SiC doping [1] is one of the most important advances since the discovery of superconductivity in this material [2]. A high $H_{irr}$ of 29 T and $H_{c2}$ of 37 T have been achieved for nano-SiC doped MgB$_2$ wires [3, 4]. The in-field $J_c$ for the nanoparticle doped samples increased by more than an order of magnitude [1, 5, 6], compared with the best results reported for un-doped samples [6-8]. This breakthrough has been confirmed by a number of groups worldwide [9-14]. In spite of intensive studies on SiC doped MgB$_2$ in the last four years, the mechanism explaining why SiC doping is special remains unclear. Moreover, it has been believed that the C substitution for boron enhances $H_{c2}$ while the defects and grain boundaries are responsible for flux pinning. The question is whether these two separate factors have a common origin. Furthermore, several tens of different dopants in $MgB_2$ have been studied thus far [3]. The results are wildly variable from case to case and some are not reproducible. The question is whether there is any common ground on the effect of many such dopants. To answer these questions, we conducted a systematic study on lattice parameters, critical temperature ($T_c$), C content, $J_c$, $H_{irr}$, and $H_{c2}$ for comparison of pure, C, and SiC doped $MgB_2$ wires. From these results we propose a unified mechanism for the enhancement in $J_c$, $H_{irr}$, and $H_{c2}$ of $MgB_2$ by nano-doping.

$MgB_2$ wires were prepared by an *in-situ* reaction method and standard powder-in-tube technique [1, 6]. Powders of magnesium (Mg, 99%) and amorphous boron (B, 99%) were well mixed for fabrication of pure $MgB_2$ wire. For processing SiC doped $MgB_2$ wire, a mixture of Mg : 2B with SiC nanoparticle powder (size 20 nm to 30 nm) with the atomic ratio of $MgB_2$ plus 10 wt% of SiC addition was prepared. These composite wires were sintered in a tube furnace at $600^oC$ to $1000^oC$ for 30min in argon atmosphere, and finally furnace-cooled to room temperature. The same procedure was used for preparation of nano-C doped $MgB_2$ wires with the nominal stoichiometric ratio of $MgB_{1.9}C_{0.1}$. All samples were characterized by X-ray diffraction (XRD). $T_c$ was defined as the onset temperature at which diamagnetic properties were observed. Transport $J_c$ of wire samples was measured using a DC method for a magnetic field range up to 16 T. The magnetoresistivity, $\rho(H,T)$, was measured with $H$ applied perpendicular to the current direction, using the four probe method in the temperature range from 4.2 K to 300 K and a field range from 0 T to 16 T. The irreversibility field, $H_{irr}$, can be deduced from $\rho(H,T)$ using the low resistivity criterion $\rho_c = 5$ nΩcm. The specific heat of the un-doped and SiC doped samples was measured using a home-made calorimeter from 2 to 45 K at zero field and at 14 T, utilizing a long relaxation technique [15].

Fig. 1 shows the *a*-axis lattice parameter, $T_c$, and actual C in the lattice versus sintering temperature for the un-doped MgB$_2$, MgB$_{1.9}$C$_{0.1}$, and 10 wt% SiC doped MgB$_2$ samples. For the un-doped sample, both *a*-axis and *c*-axis lattice parameters calculated from XRD remain constant with increasing sintering temperature. The $T_c$ shows a linear increase with increasing sintering temperature, which is attributable to the improvement in crystallinity for samples sintered at high temperature. In comparison, the *a*-axis lattice parameter shows a large drop for the SiC doped samples sintered at all three temperatures. This indicates that the C substitution for B takes place at a temperature as low as 650$^o$C. The actual C substitution level estimated from the *a*-axis change [16] shows a gradual increase with increasing sintering temperature for the SiC doped sample. The $T_c$ for the SiC doped sample is dependent on two opposite factors: C substitution level and crystallinity. An increase in C content should reduce $T_c$, while an increase in sintering temperature improves the crystallinity, and hence the $T_c$. Since there is a moderate increase in C substitution level with increasing sintering temperature, the increase in $T_c$ for SiC doped samples with increasing sintering temperature is attributable to the improvement in crystallinity. In contrast, all the C doped MgB$_2$ showed the opposite trend, i.e., there is a more pronounced decrease in the *a*-axis parameter with increasing sintering temperature, while the actual C substitution level for B increases with increasing sintering temperature, consistent with previous works [17-20]. The $T_c$ at a sintering temperature of 900$^o$C decreases notably as a result of a rapid increase in the C substitution level (Fig. 1). It is evident that the C substitution level that can be easily achieved by SiC doping with sintering at 650$^o$C requires sintering temperatures up to 900-1000$^o$C for C and carbon nanotubes (CNT) [18-20]. Thus, the unique feature of nano-SiC doping is the high reactivity of SiC, allowing it to achieve relatively high C substitution levels at low temperature. Table 1 lists the X-ray diffraction (XRD) full width at half maximum (FWHM) of the (110) peak for pure, SiC, and C doped samples sintered at different temperatures. The low-temperature processed samples and doped samples have larger values of the FWHM, indicating small grains and imperfect crystallinity.

This lower sintering temperature has a significant advantageous effect on $J_c(H)$ for un-doped and SiC doped samples, as shown in Fig. 2. It should be noted that the $J_c$ values for the SiC doped samples are significantly higher than those for the un-doped samples at both sintering temperatures. For example, the $J_c$ for the SiC doped wire at 4.2 K and 12 T is higher than that of the un-doped wire by an order of magnitude when both are sintered at 650°C while the $J_c$ for the doped wire is 40 times that of the un-doped wire when sintered at 1000°C. Furthermore, the SiC doped sample sintered at 650°C shows a clearly better $J_c$ than that sintered at 1000°C. The $J_c$ for the nano-SiC doped sample sintered at 650°C reached 10,000 Acm$^{-2}$ at 4.2 K and 12 T, which is the best in-field $J_c$ value achieved so far.

In comparison, the $J_c(H)$ for C doped MgB$_2$ wire showed an opposite trend with sintering temperature. That is, the higher sintering temperature (950°C) leads to strong improvement in $J_c$, while the lower sintering temperature results in little improvement in $J_c$ (Fig. 2). This is attributable to the fact that there is little C substitution for B at lower sintering temperature, as shown in Fig. 1. Since the C substitution level increases with increasing sintering temperature for C doped MgB$_2$, a higher sintering temperature leads to a higher level of C substitution and hence improved $J_c$. However, the higher sintering temperature also causes more grain growth, resulting in weakened flux pinning in C doped MgB$_2$.

Fig. 3 shows $H_{irr}$ as a function of the temperature for the SiC doped and un-doped samples. It should be noted that the $H_{irr}$ values derived from the magnetoresistivity are the same as those obtained from Kramer extrapolation plots, as shown by the half-filled symbols. For comparison, we plotted the $H_{c2}$ from pellet samples processed under similar conditions [21]. The SiC doping enhances both $H_{irr}$ and $H_c$. In particular, it is worth noting that the $H_{irr}$ reached 10 T at 20 K, exceeding that of NbTi at 4.2 K. The upturn in the $H_{c2}$ curve provides strong evidence for predictions from the two-gap superconductivity scenario [22].

From the above results we can propose a mechanism to explain why nano-SiC doping is so special compared to doping with all other C containing compounds. SiC is a highly stable compound. In bulk form, SiC substrates have been used to deposit $MgB_2$ thin films, which achieved a record high $H_{c2}$ [23]. However, SiC at the nanoscale becomes highly reactive. The nano-SiC reacts with Mg at a temperature as low as 600°C [6], which releases highly reactive, free C on the atomic scale as described by the reaction:

$$SiC + 2\,Mg = Mg_2Si + C \qquad (1).$$

Coincidentally, the formation reaction of $MgB_2$ from Mg and B takes place at the same temperature of 600°C. Because the free and highly reactive C is available, the C can be easily incorporated into the lattice of $MgB_2$ and substitute into B sites via the reaction:

$$Mg + (2-x)\,B + x\,C = MgB_{2-x}C_x \qquad (2).$$

This dual reaction mechanism can be well demonstrated by the reduction in the $a$-axis parameter, the decrease in $T_c$ (Fig. 1), and the occurrence of $Mg_2Si$ for a sintering temperature of 600°C. Because of the dual reactions taking place simultaneously at the same temperature, the by-products, such as $Mg_2Si$ and excess C, can be embedded within the $MgB_2$ grains as nano-inclusions. Partial substitution of C for B as a result of nano-SiC doping induces disorder on the lattice sites, which leads to the enhancement of the $H_{c2}$. At the same time, C substitution causes reduction in the grain size, as evidenced by the increase in the full width at half maximum (FWHM) in XRD in both SiC and C doped samples, and hence enhances the grain boundary pinning. The SiC doping allows an increase in the density of grain boundaries, creation of defects such as dislocations and stacking faults [2], and highly dispersed nano-inclusions within the grains [4], which can act as effective pinning centres for improving $J_c(H)$ behavior.

The XRD data indicate that the average C substitution for B in the SiC doped $MgB_2$ is about 2 at% of B. However, electron energy loss spectroscopy (EELS) analysis detected some $MgB_2$ crystals without any C peak (the inset in Fig 4). This localized C content fluctuation will lead to a fluctuation in $T_c$, and hence a broad $T_c$ distribution in the sample. The superconducting contribution to the specific heat $\Delta C_e/T = (C_e(H = 0) - C_e(14\ T))/T$ as a function of temperature was obtained for the un-doped and SiC doped samples. The superconducting transition was analysed by means of a particular deconvolution method [24] in order to determine the $T_c$ distribution in the sample volume (Fig. 4). The SiC doped sample showed a broader $T_c$ distribution than the un-doped one. The localized fluctuation in C substitution for B will cause structural distortion, as evidenced by the formation of a nanodomain structure as observed previously in SiC doped $MgB_2$ [25]. These nanodomains have a rectangular shape with a domain size of 2 to 4 nm. The domain boundaries trap numerous defects caused by the rotation of nanodomains. Transmission electron microscope (TEM) images indicate that both the grain and inclusion sizes (20 nm to 50 nm) for the sample sintered at 650°C are much smaller than for the sample sintered at 1000°C (200 nm to 400 nm). The fine grains mean more grain boundaries, which together with fine embedded inclusions can act as strong pinning centres, resulting in better $J_c$ for the sample sintered at 650°C [6].

According to the dual reaction model we can evaluate and classify a broad range of dopants into the following sequence in terms of benefits to $J_c$, $H_{irr,}$ and $H_{c2}$: the first group includes dopants such as SiC [1] and carbohydrates (CH) [26], which can have a reaction and C substitution at the same temperature as $MgB_2$ formation; the second group those such as nano-C [19], CNTs [20, 21], and $B_4C$ [11, 27, 28], which can have a reaction and C substitution at temperatures higher than that for $MgB_2$ formation; the third group those such as Si [29] and a number of silicides [16], which can react at the same temperature as $MgB_2$ formation, but without C substitution; and the fourth group those such as BN, MgO [30], etc., which have no reaction and no substitution. The last group has little positive, if not negative, effects on $J_c$, $H_{irr}$, and $H_{c2}$, even on the nano-scale.

In summary, a systematic study on the effects of sintering temperature on the lattice parameters, C content, and electromagnetic properties allows us to demonstrate a unified mechanism, according to which the optimal doping effect can be achieved when the C substitution and $MgB_2$ formation take place at the same time at low temperatures. The C substitution is responsible for the enhancement in both $H_{c2}$ and flux pinning. C substitution for B induces disorder in lattice sites, increases in resistivity, and hence enhancement in $H_{c2}$, while C substitution together with low temperature processing results in reduction in grain size, fluctuation in $T_c$, extra defects, and embedded inclusions that enhance flux pinning. SiC doping takes advantage of both C substitution and low temperature processing. An understanding of the dual reaction model has led to the discovery of the advantages of CH doping in $MgB_2$, resulting in a significant enhancement in $J_c$, $H_{irr}$, and $H_{c2}$ [26]. CHs decompose at temperatures near that of $MgB_2$ formation, thus producing highly reactive C, not dissimilar to the case of SiC doping. The model has significant ramifications with respect to the fabrication of other carbon containing compounds and composites.

The authors thank Dr. R. Klie for his assistance in EELS analysis, and Drs. E.W. Collings, H. Kumakura, T. Silver, and Mr M. Tomsic for helpful discussions. The work is supported by the Australian Research Council, Hyper Tech Research Inc, and CMS Alphatech International Ltd.

———————————

\* Electronic address: shi@uow.edu.au

**References**

1. S. X. Dou *et al*., Appl. Phys. Lett. **81**, 3419 (2002).

2. J. Nagamatsu *et al*., Nature **410**, 63 (2001).

3. S. X. Dou *et al*., IEEE Trans. Appl. Supercond. **15**, 3219 (2005).

4. M. D. Sumption *et al*, Appl. Phys. Lett. **86**, 092507 (2005).

5. S. X. Dou *et al*., J. Appl. Phys. **96**, 7549 (2004).

6. S. Soltanian *et al*., Supercond. Sci. Technol. **18**, 658 (2005).

7. R. Flükiger *et al*., Physica C, **385**, 286 (2003).

8. W. Goldacker *et al*., Supercond. Sci. Technol. **14**, 787 (2001).

9. H. Kumakura *et al*. Appl. Phys. Lett. **84**, 3669 (2004).

10. A. Serquis *et al*., Supercond. Sci. Technol. 17, L35 (2004).

11. S. Ueda *et al*., Physica C 426-431, 1225-1230 (2005).

12. W. Pachla *et al*., Supercond. Sci. Technol. **19**, 1 (2006).

13. Y. Ma *et al*., Supercond. Sci. Technol. **19**, 133 (2006).

14. E. Martinez and R. Navarro, Appl. Phys. Lett., **85**, 1383 (2004).

15. D. Sanchez *et al*., Physica C, **200**, 1 (1992).

16. S. Lee *et al*., Physica C **412-414**, 31 (2004).

17. W. K. Yeoh *et al*., Supercond. Sci. Technol. **19**, 596 (2006).

18. S. X. Dou *et al*., Appl. Phys. Lett. **83**, 4996 (2003).

19. J. H. Kim *et al*., Appl. Phys. Lett. **89**, 122510-1 (2006).

20. S. X. Dou *et al*., Adv. Mater. **18**, 785 (2006).

21. A. Serquis *et al.,* Mater. Res. Soc. Symposium Proceedings vol. EXS-3, pp. 161 (2004).

22. A. Gurevich, Phys. Rev B **67**, 184515-1 (2003).

23. V. Braccini *et al*., Phys. Rev. B **71**, 012504-1 (2005).

24. Y. Wang *et al*., Supercond. Sci. Technol., **19**, 263 (2006).

25. S. Li *et al*., Appl. Phys. Lett., **83**, 314 (2003).

26. X. L. Wang *et al*., Physica C **408-410**, 63 (2004).

27. J. H. Kim *et al*., Appl. Phys. Lett., **89**, 142505-1 (2006).

28. R. H. T. Wilke *et al*., Phys. Rev. Lett. **92**, 217003 (2004).

29. P. Lezza *et al*., Supercond. Sci. Technol. **19** 1030 (2006).

30. S.X. Dou *et al*. private communication.

Table 1. Comparison of full width of half maximum (FWHM) of (110) peak for un-doped, SiC and C doped $MgB_2$ sintered at different temperatures.

| Doping | Sintering temperature ($^oC$) | FWHM (110) ($^o$) |
|---|---|---|
| Un-doped | 650 | 0.448 |
| | 825 | 0.425 |
| | 1000 | 0.326 |
| SiC | 650 | 0.550 |
| | 825 | 0.532 |
| | 1000 | 0.410 |
| C | 700 | 0.505 |
| | 800 | 0.499 |
| | 1000 | 0.437 |

**Figure Captions**

FIG. 1 The *a*-axis lattice parameter, critical temperature ($T_c$), and actual C substitution level, x, in $Mg(B_{1-x}C_x)_2$ versus sintering temperature.

FIG. 2. The critical current density ($J_c$) at 4.2 K versus magnetic field for SiC doped and un-doped $MgB_2$ wires sintered at 650°C and 1000°C, and C doped wire sintered at 700°C and 950°C

FIG. 3. Irreversibility field ($H_{irr}$) as a function of temperature for pure and 10 wt% SiC doped $MgB_2$ and un-doped samples processed at 650°C. An $H_{c2}$ curve reported previously [21] was plotted for comparison.

FIG. 4. $T_c$ distribution of nano-SiC doped and un-doped $MgB_2$ wires derived from specific heat measurements. The inset contains an electron energy-loss spectroscopy (EELS) spectrum showing that there is no carbon in some individual grains in the SiC doped $MgB_2$, while the average C substitution level is about 2% of B.

FIG 1

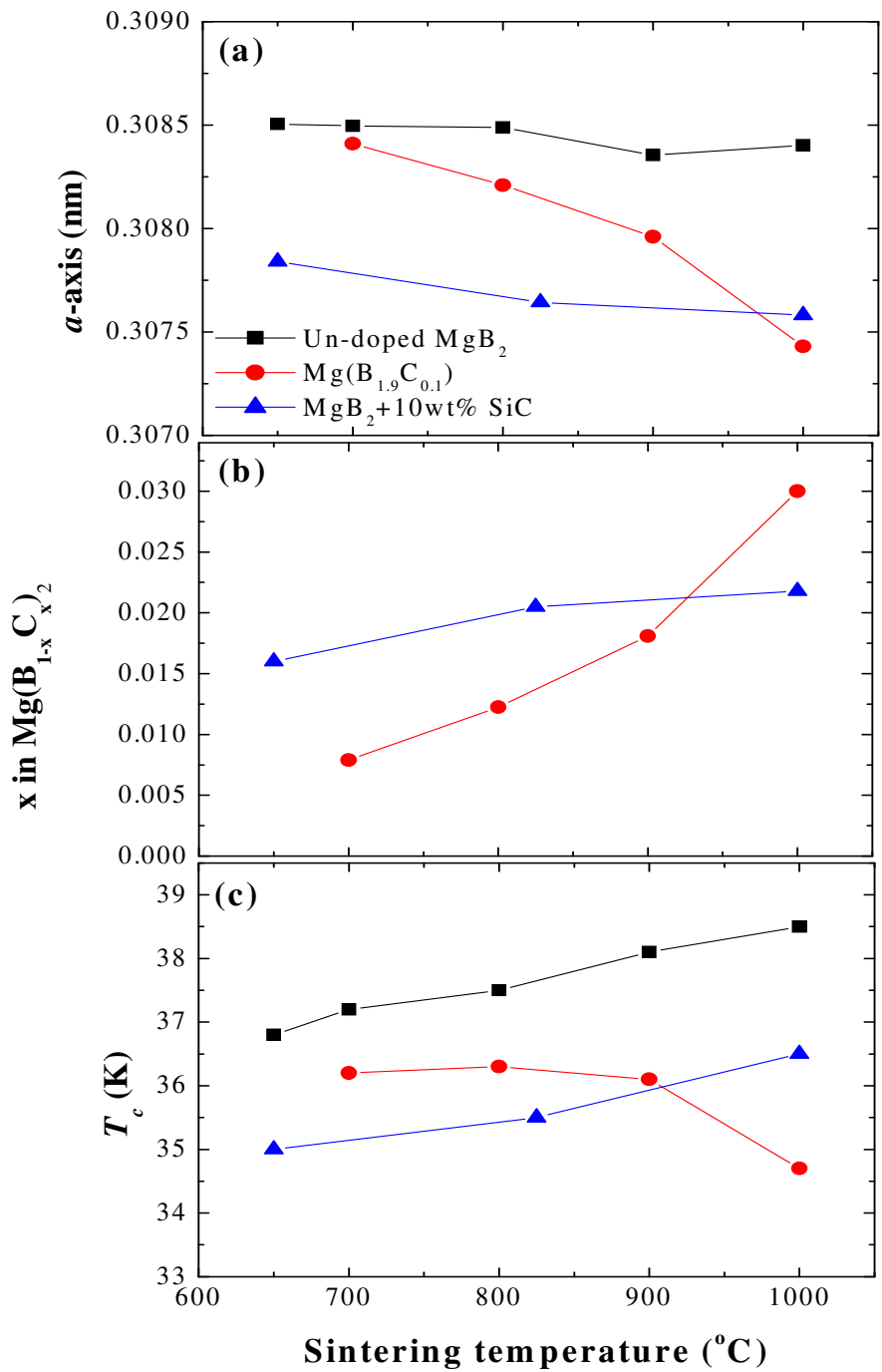

Dou *et al.*

FIG 2

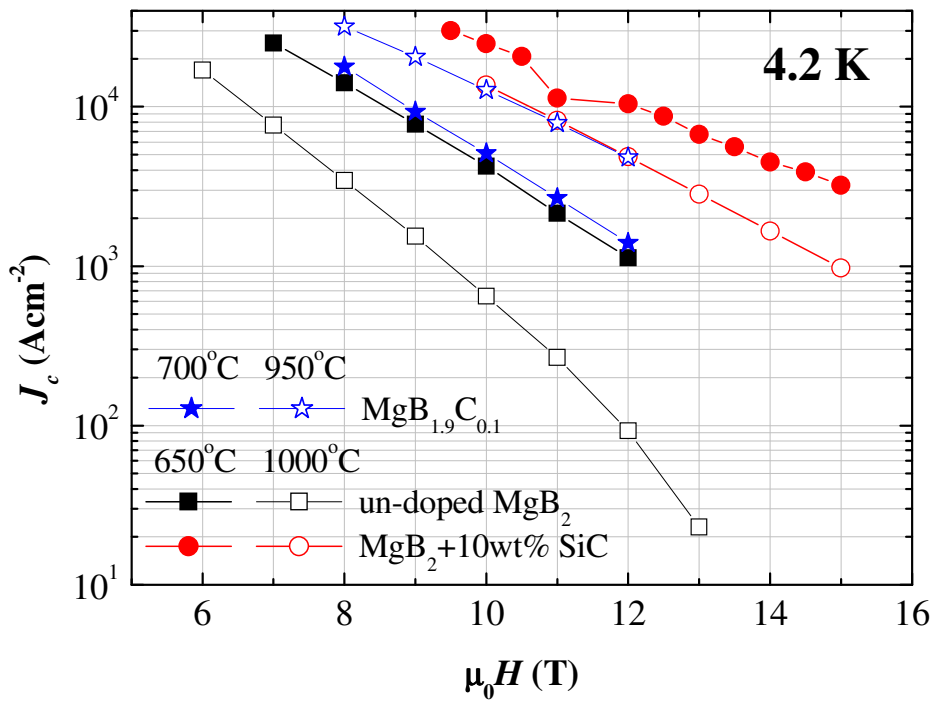

Dou *et al.*

FIG 3

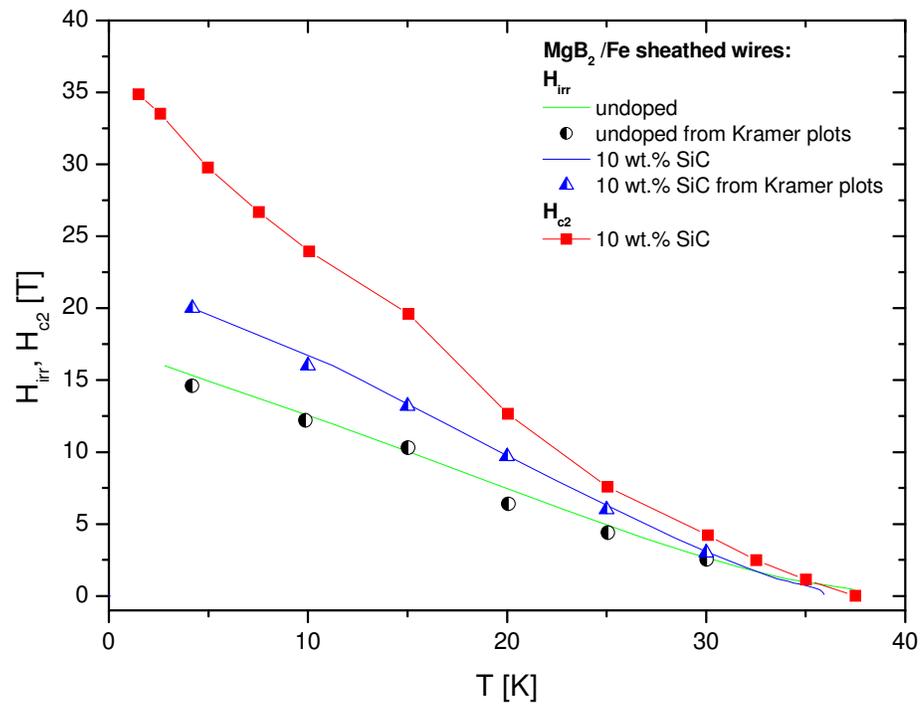

Dou *et al.*

FIG 4

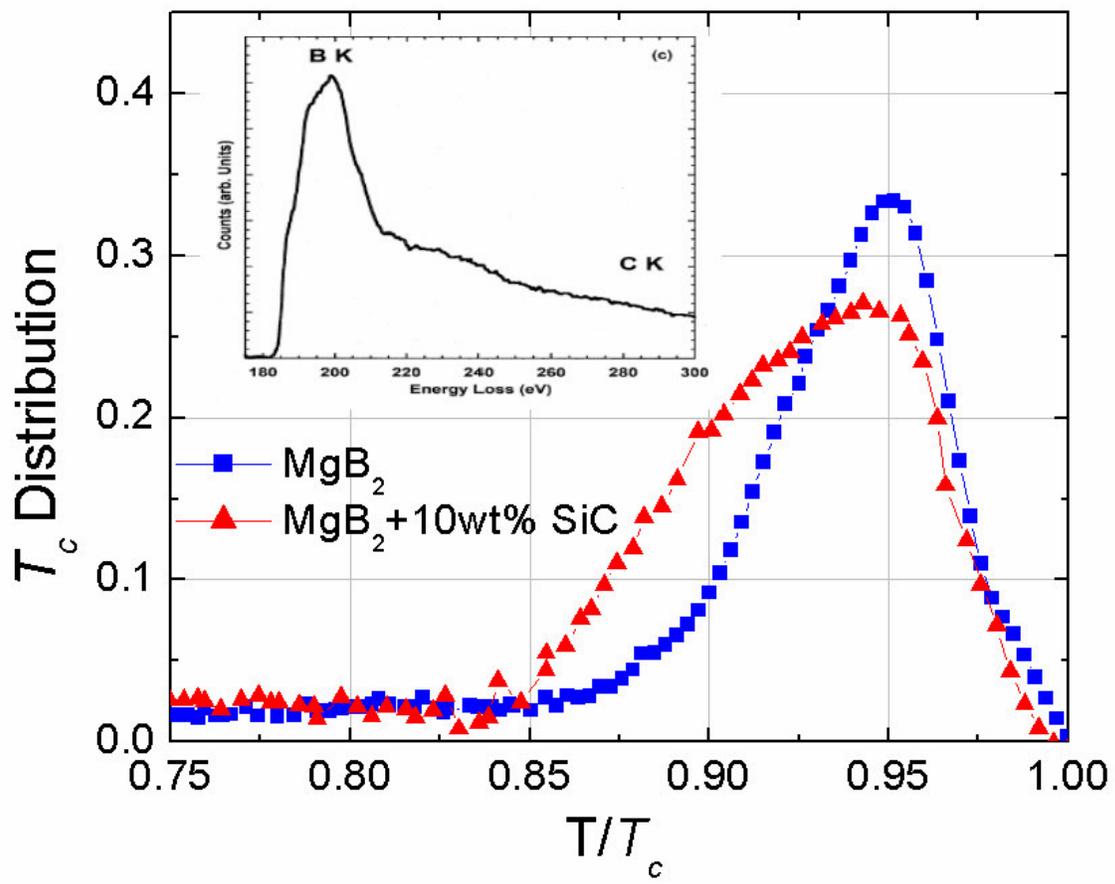